\documentclass[fleqn,twoside]{article}
\usepackage{espcrc2}
\usepackage{latexsym}
\usepackage{amssymb}
\usepackage{graphicx}
\usepackage[figuresright]{rotating}

\newcommand{\AmS}{{\protect\the\textfont2
  A\kern-.1667em\lower.5ex\hbox{M}\kern-.125emS}}
\newcommand{\be}{\begin{equation}}
\newcommand{\ee}{\end{equation}}
\newcommand{\bea}{\begin{eqnarray}}
\newcommand{\eea}{\end{eqnarray}}
\newcommand{\bean}{\begin{eqnarray*}}
\newcommand{\eean}{\end{eqnarray*}}

\title{Investigation of gauge-fixed pure $U(1)$ theory at strong
coupling \thanks{Talk presented by S. Basak at Lattice 2001}}

\author{S. Basak\address[SINP]{Theory Division, Saha Institute of
Nuclear Physics, \\ 1/AF Salt Lake, Calcutta 700 064, India}
\address{Department of Physics, NND College, \\ 170/436 N.S.C.
Bose Road, Calcutta 700 092, India}
        and
        Asit K. De \addressmark[SINP]}
       
\begin{document}

\begin{abstract}
We numerically investigate the phase diagram of pure $U(1)$ gauge
theory with gauge fixing at strong gauge coupling. The FM-FMD phase
transition, which proved useful in defining Abelian lattice chiral gauge
theory, persists also at strong gauge coupling. However, there the 
transition seems no longer to be continuous. At large gauge couplings
we find evidences for confinement. 
\end{abstract}

\maketitle

%%%%%%%%%%%%%%%%%%%%%%
\section{INTRODUCTION}
%%%%%%%%%%%%%%%%%%%%%%
Recently it has been shown \cite{bock1,bd1} that nonperturbative
gauge fixing \cite{shamir1} can be applied successfully to decouple
the longitudinal gauge degrees of freedom ({\em dof}) from Abelian
lattice chiral gauge theories ($L\chi GT$) in the limit of zero
gauge coupling.  After one gauge transforms a gauge-noninvariant
$L\chi GT$ proposal like the Smit-Swift model or the domain wall
waveguide model, one picks up the longitudinal gauge {\em dof},
the radially frozen scalars, explicitly in the action. The job of
gauge fixing is to find a phase transition where the gauge symmetry
would be restored with the scalars decoupled.

The gauge fixing proposal has also been successful in describing
pure $U(1)$ gauge theory \cite{bock2} where pure QED is recovered
at weak gauge coupling. In the present work we extend the study
of gauge-fixed pure $U(1)$ gauge theory on lattice at strong gauge
couplings. The major concern here is whether the FM-FMD
transition, where the gauge symmetry is restored and was useful
in defining $L\chi GT$, still exists at large $g$ and if so, what
is the order of the transition. The gauge-fixed pure $U(1)$
theory can be looked upon as another lattice regularization of
the strongly coupled $U(1)$ theory and its strong coupling
properties like confinement can also be investigated.

%%%%%%%%%%%%%%%%%%%%%%%%%%%%%%%%%%%%%%%%
\section{GAUGE-FIXED PURE $U(1)$ THEORY}
%%%%%%%%%%%%%%%%%%%%%%%%%%%%%%%%%%%%%%%%
The gauge-fixed pure gauge action for compact $U(1)$
\cite{shamir1,bock3}, where the ghosts are free and decoupled, is:
\be
S_B(U) = S_G(U) + S_{Gf}(U) + S_{Ct}(U) \label{action}
\ee
where $S_G$ is the usual Wilson plaquette action, $S_{Gf}$ is the
gauge fixing term and $S_{Ct}$ are appropriate counterterms given by,
\bea
S_G &=& \frac{1}{g^2} \sum_{x\,\mu <\nu} \left( 1 -
{\rm Re}\, U_{\mu\nu x} \right) \label{plact}\\
S_{Gf} &=& \tilde{\kappa} \left( \sum_{xyz} \Box(U)_{xy}
\Box (U)_{yz} - \sum_x B_x^2 \right) \nonumber\\
&& \label{gfact}\\
S_{Ct} &=& - \kappa \sum_{\mu x} \left( U_{\mu x} +
U_{\mu x}^\dagger \right) \label{ctact}
\eea
where $\Box(U)$ is the covariant lattice Laplacian with
$\tilde{\kappa}=1/(2\xi g^2)$ and
\be
B_x  = \sum_\mu \left( \frac{{\cal A}_{\mu x-\mu} +
{\cal A}_{\mu x}}{2}\right)^2,
\ee
where ${\cal A}_{\mu x} {\rm = Im}\, U_{\mu x}$. $S_{Gf}$ is not
just a naive transcription of continuum covariant gauge fixing
term it has in addition appropriate irrelevant terms. As a result,
$S_{Gf}$ has a unique absolute minimum at $U_{\mu x}= 1$,
validating weak coupling perturbation theory around $g=0$ or
$\tilde{\kappa}=\infty$ and in the naive continuum limit reduces
to $1/2\xi \int d^4x (\partial_\mu A_\mu)^2$. Validity of weak
coupling perturbation theory together with perturbative
renormalizability helps determining the form of the counter
terms to be present in $S_{Ct}$. It turns out that the most
important bosonic counterterm is the gauge field mass counterterm
given by (\ref{ctact}).

Our philosophy here has been to take a lattice theory given by
(\ref{action}) having the correct weak coupling limits, and then
try and find out the strong coupling properties of the same theory.
This is the best one can do.

%%%%%%%%%%%%%%%%%%%%%%%%%%%%%%%%%%%%%%%%%%%%%
\subsection{Phase structure at weak coupling}
One can have insight into the phase structure of the weak
coupling theory from the leading order effective potential
$V_{cl}$, obtained by perturbative expansion of $U_{\mu x}=
\exp igA_{\mu x}$ around $U_{\mu x}=1$ and then requiring
the gauge potential $A_\mu$ to be constant,
\bea
V_{cl} &=& \kappa \left[ g^2 \sum_\mu A_\mu^2 +
\cdots \right] + \nonumber\\
&& \frac{g^4}{2\xi} \left[ \left( \sum_\mu A_\mu^2 \right)
\left( \sum_\mu A_\mu^4 \right) + \cdots \right]. \label{vcl}
\eea
$\kappa\equiv\kappa_c =0$ signals a continuous phase transition
(because the gauge boson mass obviously vanishes) between two
broken phases: a ferromagnetic (FM) phase for $\kappa > 0$ and
a directional ferromagnetic (FMD) phase for $\kappa < 0$. FMD
phase is characterize by rotational noninvariance (minimum of
$V_{cl}$ is at $A_\mu \neq 0$). At the FM $\searrow$ FMD
transition the longitudinal gauge {\em dof} do not scale and
hence get decoupled. 

%%%%%%%%%%%%%%%%%%%%%%%%%%%%%%%%%%%%%%%
\section{NONPERTURBATIVE PHASE DIAGRAM}
%%%%%%%%%%%%%%%%%%%%%%%%%%%%%%%%%%%%%%%
To obtain the phase diagram of the gauge-fixed pure $U(1)$ theory,
given by the action (\ref{action}), in $(\kappa,\,\tilde{\kappa})$
-plane for fixed values of gauge coupling, we defined the following
observables:
\bea
E_P &=& \frac{1}{6L^4} \left\langle \sum_{x, \mu<\nu} {\rm Re}\,
U_{\mu \nu x} \right\rangle \label{obsep} \\
E_\kappa &=& \frac{1}{4L^4} \left\langle \sum_{x, \mu} {\rm Re}\,
U_{\mu x} \right\rangle \label{obsek} \\
V &=& \left\langle \sqrt{\frac{1}{4} \sum_\mu \left(\frac{1}{L^4}
\sum_x {\rm Im}\, U_{\mu x} \right)^2} \;\right\rangle. \label{obsv}
\eea
$E_P$ and $E_\kappa$ are not order parameters but they signal
phase transitions by sharp changes. We expect $E_\kappa \neq 0$
in the broken symmetric phase FM and FMD and $E_\kappa \sim 0$
in the symmetric (PM) phase. Besides, $E_\kappa$ is expected to have
a large slope at $2^{nd}$ order phase transition (infinite slope in
the infinite volume limit) and a discrete jump at $1^{st}$ order
transitions. The true order parameter is $V$ which allows us to
distinguish the FMD phase, where $V \neq 0$, from the other phases
where $V \sim 0$.

We also probed the pure gauge system by quenched staggered
fermions by measuring the chiral condensate,
\be
\langle \bar{\chi} \chi \rangle_{m_0} = \frac{1}{L^4} \sum_x
\langle M^{-1}_{xx} \rangle \label{chcnd}
\ee
where $M$ is the fermion matrix and $m_0$ is the bare mass.

%%%%%%%%%%%%%%%%%%%%%%%%%%%%%%
\subsection{Numerical results}
\begin{figure}[htb]
\begin{center}
\vspace{-2.0cm}
\hspace*{-1.1cm}\includegraphics[width=8.8cm,height=10.0cm]{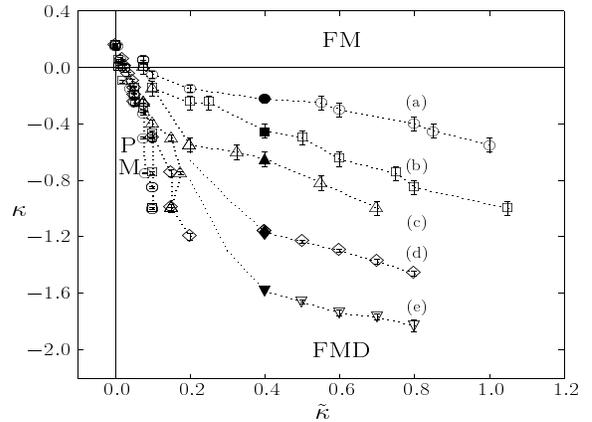}
\vspace{-4.3cm}
\caption{Phase diagram for: (a) $g=0.8\, (\bigcirc)$, (b) $g=1.0\,
(\square)$, (c) $g=1.1\, (\bigtriangleup)$, (d) $g=1.2\, (\Diamond)$,
(e) $g=1.3\, (\bigtriangledown)$ (filled symbols for $10^4$ and
empty symbols for $4^4$).}
\label{phase}
\end{center}
\vspace{-1.0cm}
\end{figure}

The Monte Carlo simulations were done with 4-hit Metropolis
algorithm on $4^4$ and $10^4$ lattices. We explored the phase
diagram in $(\kappa, \tilde{\kappa})$-plane at gauge couplings
$g\,=\,0.6,\,0.8,\,1.0,\,1.1,\,1.2,\,1.3$ and 1.4 over a range
of 0.30 to -2.30 for $\kappa$ and 0.00 to 1.00 for $\tilde{\kappa}$.
Each data point consisted of 45,000 Metropolis sweeps for $4^4$
lattice and 6,500 sweeps for $10^4$ lattice. The autocorrelation
length was less than 10 for both lattices.

Figure \ref{phase} collectively shows the phase diagram
in $(\kappa,\,\tilde{\kappa})$-plane for the different gauge
couplings. The diagram looks qualitatively the same for
all gauge couplings. At weaker gauge couplings ($g\,=\,0.6,\,0.8,
\,1.0$) the FM-FMD transition (the dotted lines roughly parallel
to the $\tilde{\kappa}$-axis) appears to be continuous, as
found previously in reduced model \cite{bock4} and as also in
the weak coupling ($g\,=\,0.6$) investigation \cite{bock2},
done previously. 

\begin{figure}[htb]
\begin{center}
\vspace{-2.0cm}
\hspace*{-1.1cm}\includegraphics[width=8.8cm,height=10.0cm]{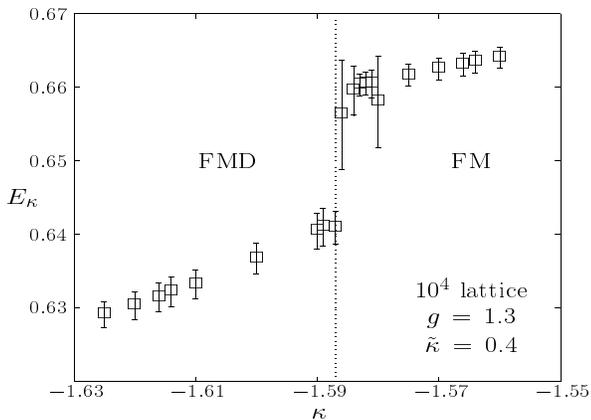}
\vspace{-4.3cm}
\caption{Discontinuity in $E_\kappa$ across FM-FMD transition,
shown in dotted line.}
\label{ordr}
\end{center}
\vspace{-1.0cm}
\end{figure}

At strong couplings ($g\,=\,1.2,\,1.3,\,1.4$) the FM-FMD
transition still exists, albeit at larger negative $\kappa$
resulting in a slimming of the FMD phase. A closer look, as
shown in fig. \ref{ordr}, at the nature of change of $E_\kappa$
across FM-FMD at these couplings indicates a discrete jump,
possibly implying a $1^{st}$ order transition.

\begin{figure}[htb]
\begin{center}
\vspace{-1.2cm}
\hspace*{-1.1cm}\includegraphics[width=8.8cm,height=10.0cm]{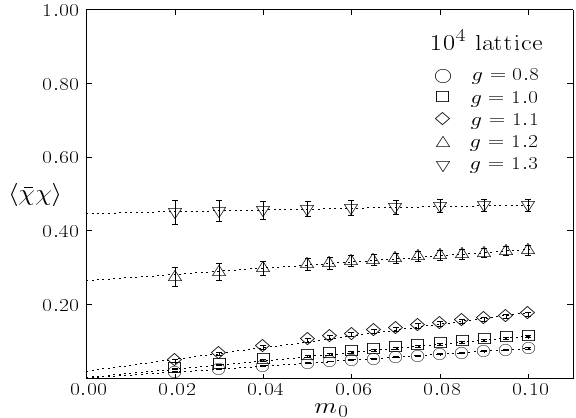}
\vspace{-4.3cm}
\caption{Quenched chiral condensate on $10^4$ lattice as a function of
$m_0$ for different $g$.}
\label{chcds}
\end{center}
\vspace{-1.0cm}
\end{figure}

Figure \ref{chcds} shows quenched chiral condensates near the
FM-FMD transition (remaining in the FM phase) for different
gauge couplings as a function of staggered fermion bare mass
$m_0$. The chiral condensates were computed with the Gaussian
noise estimator method. Anti-periodic boundary condition in
one direction was employed. Figure \ref{chcds} clearly indicates
that for weaker gauge couplings ($g < 1.1$) the chiral condensates
in the chiral limit (obtained by linear extrapolation) vanish.
However, for stronger gauge couplings ($g > 1.1$) the chiral
condensates are not zero in the chiral limit. The above observation
may signal confinement on the lattice for the strongly coupled
gauge-fixed pure $U(1)$ theory.

The gauge-fixed pure $U(1)$ theory with a dim-2 mass counterterm
has an expected weak coupling behavior persisting upto $g=1.0$.
At stronger couplings there is indication for confinement. However,
the confinement seems to be a lattice artifact because the FM-FMD
transition  probably is of $1^{st}$ order.

We thank Tilak Sinha for help with part of the simulations.

%%%%%%%%%%%%%%%%%%%%%%%%%%

\end{document}